\documentclass[a4paper,12pt]{article}
\usepackage{setspace}
\usepackage[utf8]{inputenc}
\usepackage[english]{babel}
\usepackage{amsmath}
\usepackage{graphicx}
\usepackage[round]{natbib}
\usepackage{comment}
\usepackage{bm}
\usepackage{url}
\usepackage{float}
\usepackage{varioref}
\usepackage[hang,small,bf]{caption}
\title{A {N}ew {A}lgorithm for {E}xploratory {P}rojection {P}ursuit}
\author{Mohit Dayal \thanks{Researcher, Applied Statistics and Computing Lab, Indian School of Business, Hyderabad (mohitdayal2000@gmail.com). The author would like to thank Professor Probal Chaudhuri  at the Theoretical Statistics and Mathematics Unit, Indian Statistical Institute, Kolkata for his invaluable comments and suggestions on an early draft of this paper.}}
\begin{document}
\maketitle
\begin{abstract}
In this paper, we propose a new algorithm for exploratory projection pursuit. The basis of the algorithm is the insight that previous approaches used fairly narrow definitions of interestingness / non interestingness. We argue that allowing these definitions to depend on the problem / data at hand is a more natural approach in an exploratory technique. This also allows our technique much greater applicability than the approaches extant in the literature. Complementing this insight, we propose a class of projection indices based on the spatial distribution function that can make use of such information.

Finally, with the help of real datasets, we demonstrate how a range of multivariate exploratory tasks can be addressed with our algorithm. The examples further demonstrate that the proposed indices are quite capable of focussing on the interesting structure in the data, even when this structure is otherwise hard to detect or arises from very subtle patterns.
\end{abstract}
\section{INTRODUCTION}
The importance of exploring data, especially through visualizations, before embarking on a formal analysis has long been recognised in the statistics community. However, the application of this general principle to a multivariate setting is neither easy nor straightforward, since most of the common visualizations like scatterplots or histograms are intrinsically 1, 2 or 3-dimensional. If one wishes to use these familiar tools for multidimensional data also, then some dimension reduction must be affected. Traditionally, this role has been filled in by the principal components transform.

Exploratory Projection Pursuit (EPP) is a generalization of the principal components for the purpose of data visualization. In EPP, one tries to search out those particular linear projections of the data that are deemed most interesting, with a view to later display these to the analyst. This search is aided by quantifying the interestingness of a projection of the data via a scalar function called the projection index.

While on paper the methodology seems highly promising, the central problems involved in its practical implementation, namely, the definition of interestingness and its quantification via a scalar function, seem rather insoluble. After all, ``interestingness" is a fairly diffuse notion, defying any sort of straightforward definition or easy quantification.

The first and the only practical algorithm to date, for carrying out exploratory projection pursuit was proposed in \citet{friedman1974projection}. Despite correctly identifying these central issues, no definitive answers were presented to either. Rather, they proposed the use of a projection index justified purely on the basis of their empirical observation of the human-computer interaction on the PRIM-9 computer.

Several treatments of these issues have subsequently appeared in the statistics literature. The views of \citet{huber1985projection}, \citet{friedman1987exploratory} and \citet{jones1987projection} arguably form the dominant school of thought in this field. In fact, most of the projection pursuit indices are grounded in this school. While \citet{friedman1974projection} defined a certain kind of structure as interesting and then designed an index for it, these authors have preferred to work in the reverse direction by first defining non-interestingness and then maximizing away it. The central dogma of this school of thought has been that the normal distribution is the least interesting parametric distribution. Thus, they have argued that by searching out the most non-normal projections, one can expect to be led to more interesting views.

While this has been the dominant view in the literature, it is instructive to recall alternative developments of the methodology that can accomplish the same task without making use of normality. After all, JC Gower in his discussion of \citet{jones1987projection} had pointed out that a case could equally be made for the uniform distribution or in fact any unimodal distribution as the modal uninteresting distribution. In fact, there have been at least two developments of the exploratory projection pursuit procedure using other parametric models. The first was \citet{naito1997generalized} who constructed an index based on maximizing non-ellipticity. \citet{nason2001robust}, on the other hand,  proposed the use of indices based on the t-distribution. Each of these authors noted that the respective indices were capable of uncovering similar structure in the data as the normality based ones. These developments serve to demonstrate the truth of Gower's insight.

It should thus be realized that there is nothing really holy about the use of multivariate normality in exploratory projection pursuit. The reasons behind the success of any of these indices in uncovering structure do not really derive from the specific parametric form of non-interestingness that is in use. Rather, it appears, that for most real multivariate datasets a strict parametric form like normality or ellipticity is so highly unrealistic, that simply by maximizing away from it, there is an appreciable chance of hitting upon some structure that may be present in the data.

The fact that more than one parametric model may be used to yield similar results should make us question both their necessity and utility in exploratory projection pursuit. The replacement of these artificial benchmarks by more realistic ones forms the principal motivation of this paper.

What does one mean by more realistic benchmarks? For this, we must first try and understand better the notion of interestingness. As already mentioned, interestingness is a rather diffuse and ambiguous concept. Thus, a structure that may be called interesting in one instance may not at all be so in another. Why is this so? We understand this to be really a reflection of the fact that data and its analysis occur not in a vacuum, but are meaningful only in a certain context. Thus interestingness really arises from the context, and it is this context that somehow must be made use of in defining it.

To clarify matters let us illustrate our arguments using the well-known Randu dataset.

The Randu random number generator is a linear congruential generator (LCG) dating back to the 1960s, and was in wide usage for many years. Unfortunately, it turns out that the quality of random numbers generated from it is rather poor. When successive numbers from the generator are arranged in triplets, the three dimensional structure obtained is essentially a collection of 15 parallel planes. The visualization of this special structure via projection pursuit has long been of interest.

As has been noted previously in the literature, this dataset really contains two distinct structures : a macro one, where the entire data lies inside a cube, and a micro one, namely the parallel planes. In isolation, both these structures are rather striking and may be deemed interesting. Further, both are rather non-normal and if one works with a normality-based index, it is hard to predict which will be preferred. Indeed this has been the experience.

For instance, \citet*{cook1993projection} found that the higher order terms of Friedman and Hall indices picked up on the micro structure, while neither the lower order terms nor the original indices did. Similarly, \citet{perisic2005projection} noted that the entropy index with a low bandwidth is attracted to the micro (parallel planes) structure, while with a high bandwidth it picks up the macro (cube) structure.

On the other hand, were it known to the analyst that the numbers have come from a random number generator, the macro (cube) structure would not be deemed an interesting structure at all. Thus when subject matter knowledge is brought in, the confusion of the traditional indices between these structures is superfluous! Unfortunately, this knowledge about the source of the data is unusable in the traditional context, to the detriment of the analysis.

This kind of ``filtering" out of structures based either on subject matter knowledge, or on some other criteria, is important because for real datasets, a number of projections could be very non-normal without really revealing anything interesting to the user. Further, by allowing the information available with the user to be used in data exploration, we give her greater control over the whole process. Thus she may decide on what structures may be most interesting, and then guide the search for projections towards such and away from others that are not so  in turn leading to more meaningful data exploration. This is similar to user interaction during a data tour, but automated and hopefully, easier to understand.

With the help of this insight, it easy to see that the parallel developments of projection pursuit methodology that concentrated on certain specific structures like clustering \citep{eslava1994some,bolton2003projection} or discrimination \citep{lee2005projection} as the standards for interestingness can ever only be half-measures. In fact, the clustering and discrimination indices are really, in some sense, a mirror image of the normality based indices. While the latter have concentrated on measuring departures from a strict parametric form, the former try to measure conformance to a special structure. Neither is capable of making use of a context-dependent information, and thus neither is really general.

In this paper, we propose a new algorithm for exploratory projection pursuit that can accomplish this task. In particular, we leave questions of interestingness and non interestingness to the user who is best placed to answer such questions. This makes the whole process more transparent and more generally applicable. The incorporation and use of such information necessitates a modification to the basic algorithm for exploratory projection pursuit.  Next, we propose the use of a class of projection indices that can make use of such information and search out interesting projections. These indices possess many of the optimality properties that have been proposed in the literature. We further suggest how benchmarks of interestingness can be arrived at even in the absence of subject matter knowledge, while yet not making use of any parametric models or asymptotic arguments. Taken together, these developments constitute a coherent alternative framework for exploratory projection pursuit that is more generally applicable.

The rest of this paper is organized as follows. In section 2, we start by recalling the traditional projection pursuit algorithm and point out its shortcomings. We use the lessons learnt to design a new algorithm that can overcome these limitations. In section 3, we propose a class of projection indices that can work with the new method. In section 4, we treat three real datasets to demonstrate how our methodology works. Two of the three examples have previously been treated in the projection pursuit literature and this affords a comparison between the existing procedures and ours. Section 5 concludes.

\section{THE ALGORITHM}
It should clear from the preceding discussion that we aim to incorporate flexible definitions of interestingness into the exploratory projection pursuit procedure. To begin, let us try to understand exactly why the usual exploratory projection pursuit procedure cannot accommodate such information. For this, it is helpful to break down the procedure into the following three steps.

\begin{enumerate}
	\item[Step 1:] The user chooses a target dimension $d$ in which the data will be visualized.
	\item[Step 2:] The projection index $p(\mathbf{A},\mathbf{X})$ is defined over all $d$-dimensional linear projections $\mathbf{A}$ of the dataset $\mathbf{X}$.
	\item[Step 3:] The index $p(\mathbf{A},\mathbf{X})$ is maximized over all possible choices of $\mathbf{A}$, with suitable constraints, if any.
\end{enumerate}

At the final stage, one projects the data onto the $\mathbf{A}$'s that correspond to substantial local optima of the projection index $p(\mathbf{A},\mathbf{X})$ for a visual examination.

It can be seen that in this procedure there is no explicit specification of what constitutes an interesting and / or uninteresting projection. In fact, this specification is implicit in the choice of the projection index. Two broad classes of such indices are available : those based on test statistics for normality, and those designed for specialized tasks like clustering and discriminant analysis. Apart from the choice of index, the usual algorithm provides no scope at all for user intervention.

For the direct incorporation of auxiliary information into projection pursuit, we propose the following modified algorithm.

\begin{enumerate}
	\item[Step 1:] The user chooses a target dimension $d$ in which the data will be visualized.
	\item[Step 2:] The user chooses a benchmark $Y$ of interestingness or non interestingness.
	\item[Step 3:] The projection index $p(\mathbf{A},\mathbf{X},Y)$ is defined over all d-dimensional linear projections $\mathbf{A}$ of the data $\mathbf{X}$ and benchmark $Y$.
	\item[Step 4:] The index $p(\mathbf{A},\mathbf{X},Y)$ is maximized over all possible choices of $\mathbf{A}$, with suitable constraints, if any.
\end{enumerate}

The essential feature of our algorithm is that it unlinks the question of interestingness or its absence from the choice of the projection index, allowing each to be dealt with in an independent manner. In other words, the notion of what is interesting or not can change from problem to problem, at the user's discretion, without necessarily requiring a change of projection index. Thus subject matter or other inputs can be meaningfully incorporated without being overwhelming for the user.

What should be the nature of the benchmark and how does one get hold of it? In our implementation of this algorithm, we require the specification of the benchmark via a dataset of the same dimensionality as the original dataset, though the number of observations can differ between the two. Essentially, we intend to explore the interesting structures in the dataset of interest in view of the information provided by the auxiliary dataset. This is accomplished by carrying out a search for either the similarities or the dissimilarities between them. Thus such a search supports the metaphor of a ``filtering" out of structures.

Though not essential to the algorithm, we suggest in the following three ways that the user may build up such benchmarks. Examples of each of these occur in the examples section.

In the first instance, the investigator may choose to use a dataset resulting from a previous study of similar nature as a benchmark, if such is available. In such a case, the benchmark represents the available subject matter knowledge. Problems could arise when new variables have been recorded, and old ones dropped. However a search for projections using the common variables may still prove very enlightening. These solution projections, expanded to the dimensionality of the new data, could be used as starting points to explore and analyse the data further.

In cases where subject matter knowledge is either not available, or one does wish to use it, there are two courses of action available. In the first, one can use factor variables that may be present in the data to partition the data into two and arrive at a benchmark / dataset pair. This can be useful where there is a comparatively small number of factor levels available and when they are easily interpretable. This is a case of exploiting the structure of the data to build up benchmarks. When the solution projections exhibit a classification structure, the technique has some of the flavour of the exploratory projection pursuit discriminant analysis of \citet{lee2005projection}. However, since we do not impose class separation as an explicit objective, this may not always happen and other kinds of structures could possibly be exposed. Thus while our technique is more general than that of \citet{lee2005projection}, it would be less powerful when one is explicitly looking for a discrimination structure.

When neither of the above options is available, there is another way to create ``objective" internal benchmarks from the data. These benchmarks are similar in spirit to the parametric models that have traditionally found use in exploratory projection pursuit. The intuition behind them is fairly simple. Under the hypothesis that the data indeed contains some special structure of a multivariate nature, then it naturally would be the result of the joint distribution of the variables. Consider then another dataset where the marginal distribution of each variable is still the same, but the joint distribution is one of independence. Such a dataset is in some sense closest to the original data, but yet does not exhibit the special structure. It is hoped that comparisons of this ``synthetic" dataset to the original one would reveal structures in the original data. Since such a synthetic dataset is closer to the original data than any parametric model, one can expect more meaningful results. Such ``synthetic" datasets can be constructed ad infinitum using the permutation principle.

In fact, such permutation benchmarks have wider utility. For example, when there are several class labels, this method is more practicable than the earlier one one which may require several binary splits. By destroying the joint distribution, one can reasonably hope that those projections that clearly show the groups in the data would be most dissimilar to this homogenised synthetic. Similar remarks apply when the goal is to obtain a clustering structure. They are also useful when attempting to uncover structure in very highly structured datasets for which it may difficult to obtain benchmarks otherwise.

Of course, other reasons can also be used to motivate the choice of a benchmark and more than one benchmark may be utilized during the course of an investigation.

\section{PROJECTION INDEX}
The second important component of our algorithm is a projection index that can quantify the extent of similarity or otherwise between particular projections of two multivariate datasets. The question of similarity is of course, guided by the type of difference that one expects to find. In an exploratory analysis, it would be prudent to ensure that the function chosen respond to any and all kinds of differences between the datasets. Since a difference in distribution is the most fundamental kind of difference we were led to consider this class of test statistics to form the basis for the projection index.

A number of such multivariate distributional tests have been proposed in the literature and in principle, any of these could be adapted to form a projection index. The best known is of course the $\chi^2$ statistic. However it is also singularly unsuitable for such use, because of the requirement of a cell partition scheme. The power of the statistic depends critically on the particular scheme used, yet as the projections change, the appearance of the point cloud can change dramatically, causing problems for any cell partition scheme. Other choices are the tree-based generalizations of the Wald Wolfowitz and Kolmogorov Smirnov statistics suggested by \citet{friedman1979multivariate}, the multivariate versions of the Kolmogorov Smirnov statistic \citep*{fasano1987multidimensional,justel1997multivariate}, and the Cramer von Mises statistic. However, we found these statistics to be rather unsuitable to serve as a basis for a projection index. We have thus chosen to base our projection index on the spatial distribution function, sometimes also called the M-distribution function \citep{koltchinskii1997m,serfling2002quantile}. This is nothing but the inverse function of the better-known spatial quantiles \citep{chaudhuri1996geometric}.

For a vector-valued random variable $\mathbf{X}$ having an absolutely continuous distribution in $\Re^d$ and $\mathbf{t}$ $\in \Re^d$, the spatial distribution function $\mathbf{G}_\mathbf{X}(\mathbf{t})$ is defined as
\begin{equation}
\mathbf{G}_\mathbf{X}(\mathbf{t}) = E{\frac{\mathbf{X}-\mathbf{t}}{\|\mathbf{X}-\mathbf{t}\|}}
\end{equation}

$\mathbf{G}_\mathbf{X}(\mathbf{t})$ can thought of as a generalization of the cumulative distribution function to multidimension. Just like the cumulative distribution function, $\mathbf{G}_\mathbf{X}(\mathbf{t})$ also characterizes the distribution of $\mathbf{X}$. However, it has the advantage of being a continuous function of its argument, making it easier to handle.

Test statistics for differences in distribution can be easily constructed using $\mathbf{G}_\mathbf{X}(\mathbf{t})$. For the the two-sample case that is of interest, we propose use of the function,
\begin{equation} \label{eq:integral}
\int_{t \in S} || \mathbf{G}_\mathbf{X}(\mathbf{t}) - \mathbf{G}_\mathbf{Y}(\mathbf{t}) || dt
\end{equation}

This statistic is similar to the Lorenz curve type measures of equality.

For the estimation of this quantity, we simple use the plug-in estimators for $\mathbf{G}_\mathbf{X}(\mathbf{t})$ and $\mathbf{G}_\mathbf{Y}(\mathbf{t})$ and evaluate the integral numerically. Further, we need to restrict the region of integration to some finite region, say $S$ for the computation of this integral. A convenient framework emerges if we choose $S$ as the region, 
$S(k) = \{ \mathbf{t} : |\bm{\mu} - \mathbf{t}| < kr \}$, 
where $\bm{\mu}$ is any estimate of location for the two datasets combined, $r > 0$ is a radius parameter and $k > 0$ serves as a multiplier. Thus we compute the integral inside a circle of radius $kr$ with centre at $\bm{\mu}$. For an estimate of $\bm{\mu}$, we use the spatial median since it fits naturally into the framework. For the radius parameter $r$, we took the distance of the farthest data point from $\bm{\mu}$. Thus, $k$ is the only parameter that needs to be set by the user \citep{perisic2005projection}. In our experience, values in the range (0.5,3) work well, and the index is not too sensitive to the choice.

The function chosen by us is preferable to the other statistics on at least three grounds : fast computation, rotational invariance and easy generalizability to higher dimensions. These have been widely recognised in the literature as requirements of a good projection index. That our index possesses the last two properties should be easily apparent, as should be the fact that none of the other statistics mentioned possess them. The question of fast computation however, deserves further comment.

It is should be easy to see that since the projection index is computed repeatedly over hundreds of candidate planes, an index that takes time to compute can slow down the whole process to a crawl. There are two moderating factors, however, on this requirement.

The first is that in the search phase, one does not need a very high degree of accuracy. Thus, if some approximation to the test statistic is available, then it can be used in the search phase to speed up matters. Once a final plane is output by the search step, an exact value can be calculated if required.

The second relaxation comes in from the kind of computer hardware available today. Multi-core CPUs have now become common in desktop computers and it is naturally useful to consider statistics that can take advantage of this processing power. In particular, statistics that are “embarrassingly parallel” to compute are preferable to those that require substantial adaptation to take advantage of parallel computation, which in turn are preferable to statistics that are essentially “single-threaded”.

In the use of our index, we exploit both these relaxations. Since the function $G_X(t)$ is continuous, we compute the integral using quasi-Monte-Carlo. Not only do we get better accuracy for a fixed number of evaluation, but it also provides a systematic way to scale up the accuracy of the approximation. Further, it also ensures a kind of quasi-stability in the index approximation, which is important during the search phase. Finally, since the functional values can be calculated at any value of $t$ independently of the functional value at other points, the evaluation of our index is an ``embarrassingly parallel" problem.

On the other hand, the other statistics like those of \citet{friedman1979multivariate} which require the computation of the MST or the KS-type statistics of \citep*{fasano1987multidimensional,justel1997multivariate} all involve discrete functions. Thus a complete evaluation of the test statistic is required each time. They all also involve some kind of sorting-like operation which are not as easily parallelized.
\section{EXAMPLES}
We implemented this algorithm on the computer in the {\tt R} language \citep*{rlang}. The programs have three largely separate components. First, the optimization relating to the calculation of the spatial median. Next, the evaluation of the integral in the projection index. And finally, the optimization of the projection index itself.

Optimizations relating to the spatial median were performed using the {\tt trust} package \citep{trust}. Evaluation of the integral was performed via quasi-Monte Carlo using Sobol sequences generated using the {\tt randtoolbox} package \citep*{randtool}. The projection index was optimized using either simulated annealing ({\tt optim} function in {\tt R}) or the geodesic optimization functions of the {\tt tourr} package \citep*{wickhamtourr2}. Throughout, a number of supporting linear algebra functions were adapted for use from this latter package.
\subsection{Randu}
The Randu problem was already encountered in the introduction. Since we wish to illustrate the creation and use of a benchmark derived from subject matter knowledge, we shall work on the assumption that it is known that the numbers have come from a random number generator. For the data itself, we turn to the {\tt datasets} package of {\tt R} which contains a fragment of the {\tt RANDU} function of {\tt VAX FORTRAN} running under {\tt VMS 1.5}.

The first step in the proposed algorithm is the specification of a benchmark dataset. The choice is rarely unique. For example in the present case, given that we are dealing with pseudo-random numbers from an Linear Congruential Generator(LCG), several benchmarks could be suggested. One could use true random numbers generated from a physical process, random numbers from sophisticated random number generators like the Mersenne-Twister or from another LCG. However, it is always best to use a benchmark dataset as closely related to the original one as possible, since such a choice concentrates attention on the features of the dataset of interest. This led us to the \citet{park1988random} MINSTD LCG, which is generally considered to be the barest minimum acceptable as a random number generator. Even this is not sufficient for uniqueness, since different random seeds for the MINSTD generator could lead one to different views. In most cases, running the algorithm with 2 or 3 different seeds seems sufficient.

\begin{figure}[H]
\centering
\includegraphics[scale=0.6]{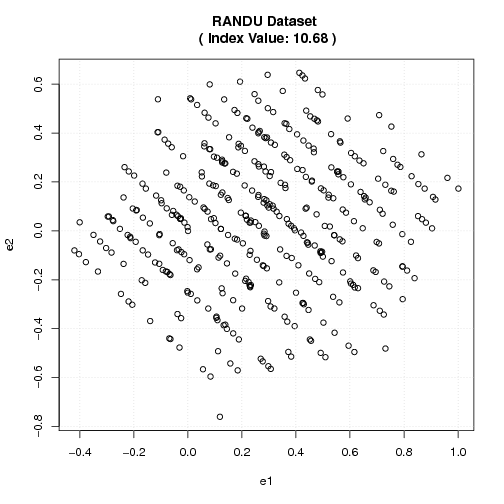}
\caption{Most interesting solution projection of the RANDU data found with parameter k = 1 and using a 50 point approximation to the integral.}
\label{fig:Randu_new3}
\end{figure}

We used simulated annealing to optimize the index. The search was initialized from 10 random projection matrices and 200 iterations were allowed from each start. The radius multiplier \textit{k} was set at 1. The best projection is given in figure \vref{fig:Randu_new3}. Five of the 10 solution projections showed the structure as clearly, though the other 4 had lower index values. The rest of the projections revealed subtle differences between the MINSTD and Randu generators.

\subsection{Colon Cancer}
Class labels often form a part of the data, with the primary question of interest centring on their influence. At other times, such labels are available, though it is not clear what bearing they have on the measurement variables. In either case, it is invaluable to apply an exploratory method to investigate the issue as a first step. In this example, we illustrate how one could accomplish this using the present algorithm.

The data used relates to colon cancer and first appeared in \citet{alon1999broad}. It relates to 22 normal and 40 tumour colon tissues and the intensities of expression of 2000 genes observed across both samples. Thus the dataset is quite high dimensional. While it would definitely be useful to build a classifier for this data, \citet{alon1999broad} raised another important issue in their paper. They argued that the separation of the two kinds of tissue resulted from the influence of \textit{all}, or at least the vast majority of genes, i.e.\ it was not attributable solely to those genes that were individually most helpful in separation. In their study, they identified genes most helpful in separation by t-ratios.

\begin{figure}[H]
\centering
\includegraphics[scale=0.6]{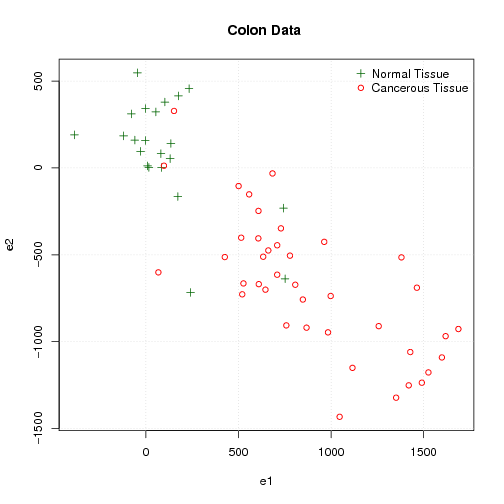}
\caption{Solution projection of the colon data found with parameter k = 2 and using a 50 point approximation. Search phase projection index 23.43. 150 point approximation 22.67, 5000 point approximation 22.68.}
\label{fig:colon_t1uncorrected}
\end{figure}

Let us start with the issue of discrimination. As always, we need a benchmark dataset. In this case it is particularly simple, since comparisons can be relative to either of the two classes, normal or cancerous tissue. This also demonstrates a case somewhat different from the usual, since we are interested in both the datasets, and further we plot both on the same graph.

\begin{figure}[H]
\centering
\includegraphics[scale=0.6]{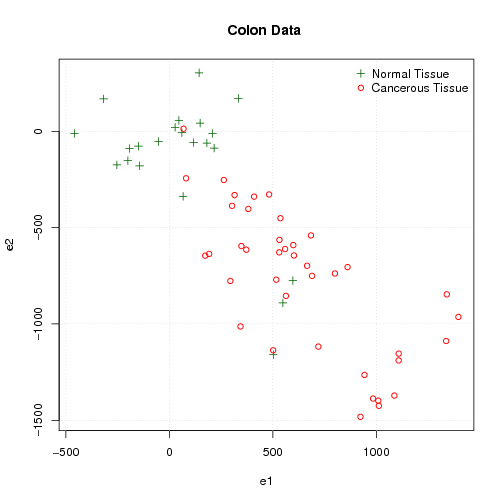}
\caption{A solution projection of the colon data showing the two group-like structure in the cancerous tissue. Found with parameter k = 2 and a 50-point approximation. Search phase projection index 22.51. 150 point approximation 21.79, 5000 point approximation 21.63.}
\label{fig:colon_t2uncorrected}
\end{figure}

Solution projections found are shown in figures \ref{fig:colon_t1uncorrected} and \ref{fig:colon_t2uncorrected}. Separation between the two kinds of tissue samples is achieved. Further, it appears from some solution projections that there may two groups in the tumour samples, one closer to the normal tissues than the other. This is the kind of insight that a visual analysis like exploratory projection pursuit can provide over an analytical one. For instance, it seems to have gone unnoticed in the original article.

Coming next to the question of whether the separation is the result of only a few genes or the vast majority of them, one first needs to identify those particular genes that could be thought of contributing most to the separation. Though one could use t-ratios, there is a more natural choice in the present context.

\begin{figure}[H]
\centering
\includegraphics[scale=0.6]{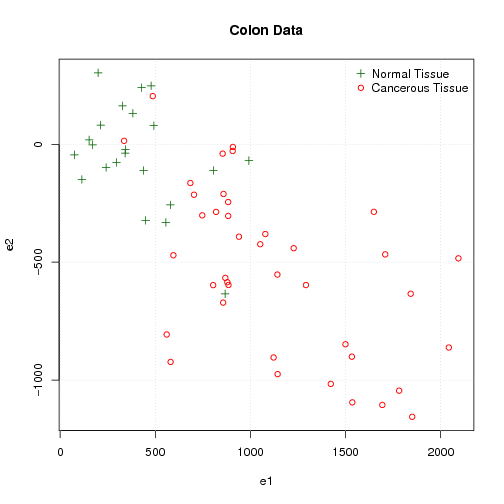}
\caption{From the projection matrix corresponding to figure \ref{fig:colon_t1uncorrected}, only the 734 rows (genes) with a norm greater than $1/\sqrt1000$ were retained. The structure is still largely the same.}
\label{fig:colon_t1uncorrectedbig}
\end{figure}

The final output of the exploratory projection pursuit is a semi-orthogonal projection matrix, whose entries one may be tempted to use for ranking genes in importance of their contribution to separation. In the present case, these matrices are of order 2000 x 2, with each column a unit vector, and each row corresponding to a gene. Under the hypothesis that the matrix entries are taken at random, each entry would have on average a magnitude of $1/\sqrt{2000}$, and each row an expected norm of $1/\sqrt{1000}$. Thus the extent to which the norm of each row deviates, either above or below this value could be thought of as indicating the importance of the corresponding gene in the projection.

\begin{figure}[H]
\centering
\includegraphics[scale=0.6]{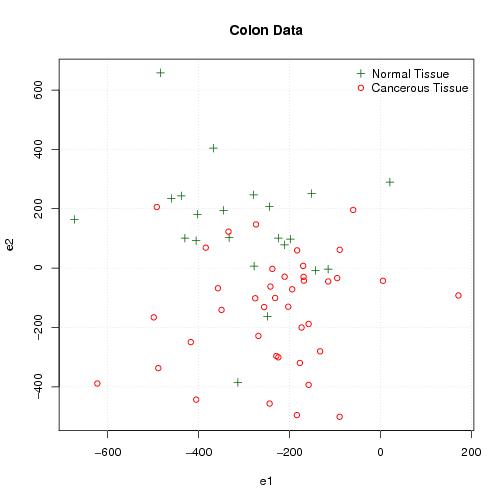}
\caption{From the projection matrix corresponding to figure \ref{fig:colon_t1uncorrected}, 1266 rows (genes) with a norm less than $1/\sqrt1000$ were retained. The structure has changed quite a bit, but the separation is still intact, albeit with higher misclassification.}
\label{fig:colon_t1uncorrectedsmall}
\end{figure}

Once the ``important" genes have been identified, and separated from the others, one course of action would be to re-run the exploratory projection pursuit on each of these two sets separately. However, one can exploit the linearity of projection pursuit to build up a quick check as follows. From each solution projection matrix, say P of order 2000 x 2 we construct two matrices L of order ${n_1~\textsc{x}~2}$ and H of order ${(2000-n_1)~\textsc{x}~2}$. The matrix L is obtained from P by dropping those rows whose euclidean norm is less than $1/\sqrt{1000}$, and H is obtained by reversing the criterion. Our procedure would lend support to \citet{alon1999broad}'s results if the separation achieved in the projections corresponding to these matrices is comparable to that obtained using P.

We reproduce the figures that result from applying this procedure on the first solution projection depicted earlier. The structure obtained from the original projection is very well preserved in figure \ref{fig:colon_t1uncorrectedbig}, which corresponds to the matrix obtained by retaining the 734 ``big" rows of the original projection matrix. However, even in figure \ref{fig:colon_t1uncorrectedsmall}, which is obtained from the 1266 ``small" rows, a separating plane can still be drawn. However, in both figures \ref{fig:colon_t1uncorrectedbig} and \ref{fig:colon_t1uncorrectedsmall} the misclassification rate has increased, something that should not be of very great concern given that neither projection is optimal in any sense. This encouraging result could be taken as an indication that the present algorithm truly picks up the subtle and fundamental patterns in the data.

\subsection{Olives Data}
This dataset relates to olive oils produced from different regions of Italy and the percentage of 8 fatty acids contained therein. One hopes to identify the production region of the different oils by studying their fatty acid composition. The identification of the production regions itself is hierarchical : they are identified both by the broad region in Italy (North, South and the island of Sardinia) and the particular collection area. Three oils come from the north: East and West Liguria and Umbria, four from the south: Calabria, North and South Apulia and Sicily, and two from the island of Sardinia : the Coast and Inland. In the present section we shall assume that none of the class labels are available and attempt to arrive at the clustering structure via the use of a permutation benchmark.

We recommend a three step strategy to identify the clusters from the solution projections. First, one obtains several solution projections from the projection pursuit algorithm. Next, these are plotted in linked displays, using for example, the {\texttt iplots} package \citep*{iplots}. Finally, brushing of the isolated points in each projection can lead to a rapid and comprehensive understanding of the data structure. The strategy is similar to one used with dynamic graphics, but by use of static displays has the advantage of being more easily understood.

\begin{figure}[H]
\centering
\includegraphics[scale=0.6]{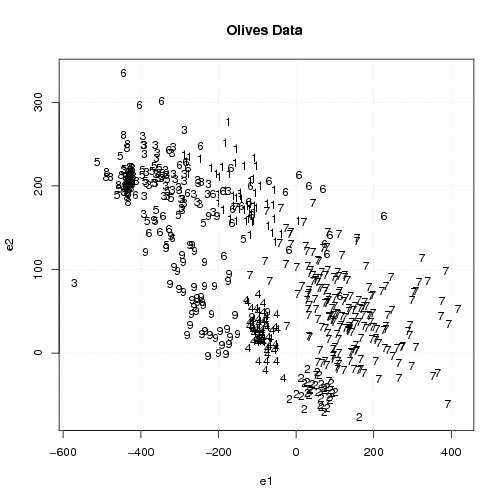}
\caption{Solution Projection with index 27.6. Clusters are well separated.}
\label{fig:olives_full_4A}
\end{figure}

\begin{figure}[H]
\centering
\includegraphics[scale=0.6]{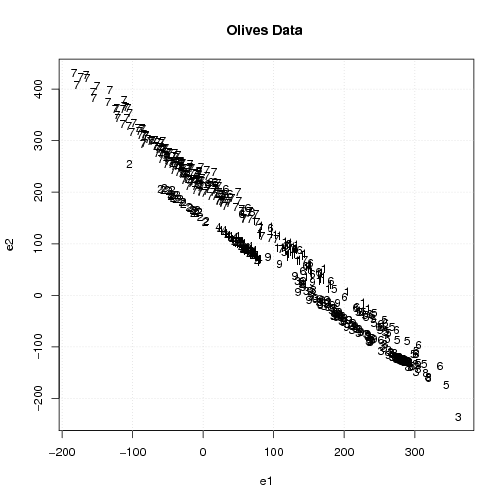}
\caption{Solution Projection with index 29.5. Shows a linear relation and clusters.}
\label{fig:olives_full_20}
\end{figure}

The figures show some typical solutions found by the projection pursuit algorithm. Qualitatively, most of the solution projections are similar, in that they display the same set of clusters, with varying degree of separation. This shows that the algorithm has been able to pick up on the clustering structure. One can see that the two oils from the island of Sardinia (symbols 2 and 4) have broken apart cleanly. Those from South Apulia (symbol 7) and West Liguria (symbol 9) also break away into separate clusters. It is not readily apparent, but the Umbrian oils (symbol 8) have broken away too, but they are so tightly clustered that they appear as smudges of ink in the top left of figure \ref{fig:olives_full_4A} and the bottom right of figure \ref{fig:olives_full_20}.

Following an isolation strategy, we removed the oils corresponding to these 5 regions from the dataset, and ran the programs with this reduced set. Interestingly, none of the two-dimensional projections showed much structure. We thus shifted to three-dimensional projections in the hope of achieving a better structure. Most of these projections showed similar structure with oils from Calabria, East Liguria and North Apulia separating out to a greater or lesser extent in different projections, while the oils from Sicily did not localize in any projection.

\begin{figure}[H]
\centering
\includegraphics[scale=0.5]{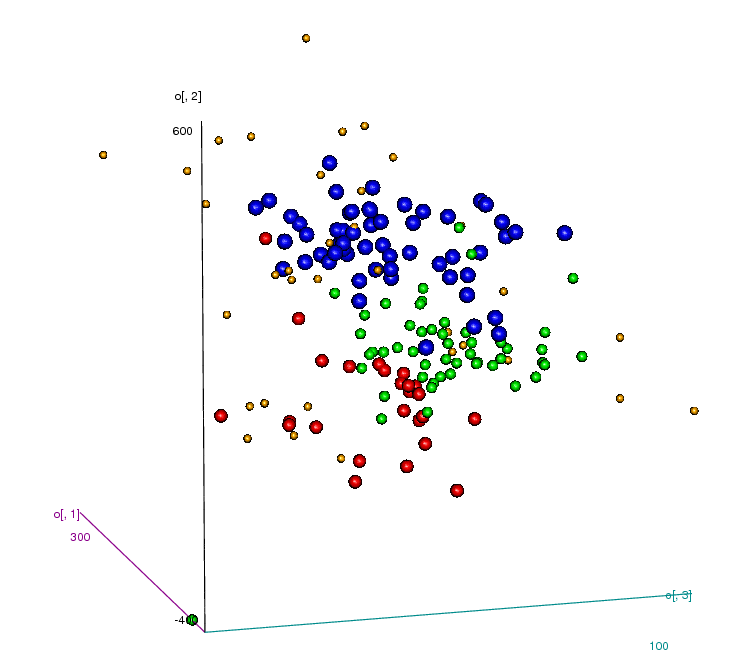}
\caption{Three Dimensional Solution Projection with index 2.54. Oils from Calabria (top), East Liguria (middle) and North Apulia (bottom) separate out, while those from Sicily (small spheres) spread out forming a sheath-like structure around this main structure.}
\label{fig:olives_3d}
\end{figure}

\section{DISCUSSION}
Methods for high dimensional data are per force highly computational in nature. While their computational nature is precisely what makes the methods feasible, there is also always the black box risk : that the algorithm used is so inflexible that there is no way for the user to either modify it or interact with its output. For exploratory methods this risk is even greater, simply because the demands for user feedback and control are so much higher. While no method can currently provide the same level of user input and interaction that are so naturally and easily incorporated in the low dimensional visualizations like scatterplots, it is still a goal worth aspiring for. On the other hand, there is also always the risk of excess. For instance, it is not uncommon for users to feel disoriented and confused when presented with dynamic graphics like data tours that are largely user-driven.

Exploratory projection pursuit has the potential to strike a balance between these two extremes: while it is highly computational, the output, which is usually a scatterplot is invariably easily understood. Unfortunately, however, the traditional algorithm for the technique is a near-perfect black box, in that it offers no scope for interaction with either the algorithm or its output. While guided tours \citep{cook1995grand} offer one way for the user to interact with the output, the more important aspect regarding selection of solution projections has remained impenetrable for the average user. The proposed algorithm thus represents an attempt at increasing the level of user feedback with the whole process, in hope that it will lead to more relevant and meaningful results from the analysis.

\scriptsize
\begin{twocolumn}
\singlespacing
\bibliographystyle{plainnat}
\bibliography{pp}
\end{twocolumn}
\end{document}